\documentclass[a4paper,twoside,twocolumn,english,showpacs,PRA]{revtex4}
\usepackage{times}
\usepackage[OT1]{fontenc}
\usepackage[latin1]{inputenc}
\usepackage{amsmath}
\usepackage{graphicx}
\usepackage{amssymb}

\makeatletter

\usepackage{babel}
\makeatother
\begin{document}

\title{Magnetic Richtmyer-Meshkov instability 
in a two-component Bose-Einstein condensate}

\author{A. Bezett}
\author{V. Bychkov}
\author{E. Lundh}
\author{D. Kobyakov}
\author{M. Marklund}

\affiliation{Department of Physics, Ume\r{a} University, SE-901 87 Ume\r{a}}
\date{\today}

\newcommand{\etal}{\emph{et al.}}
\newcommand{\rC}{\text{\bf{C}}} 
\newcommand{\rI}{\text{\bf{I}}} 
\newcommand{\PC}{\mathcal{P}_{\rC}}
\newcommand{\bef}{\hat{\psi}}
\newcommand{\bfI}{\bef_{\rI}}
\newcommand{\cf}{\psi_{\rC}}
\newcommand{\cfp}{\psi_{\rC'}}
\newcommand{\ecut}{\epsilon_{\rm cut}}
\newcommand{\CF}{c-field}
\newcommand{\Nc}{N_{\rm{cond}}}
\newcommand{\psic}{\psi_{\rm{cond}}}
\newcommand{\ac}{\alpha_{\rm{cond}}}
\newcommand{\thold}{t_{\rm{hold}}}
\newcommand{\nmin}{n_{\min}}
\def\x{\mathbf{r}}
\newcommand{\xa}{(\x)}

\begin{abstract}
The magnetically induced Richtmyer-Meshkov instability in a two-component Bose-Einstein condensate is investigated. We construct and study analytical models describing the development of the instability at both the linear and nonlinear stages. The models indicate new features of the instability: the influence of quantum capillary waves and the separation of droplets, which are qualitatively different from the classical case. We perform numerical simulations of the instability in a trapped Bose-Einstein condensate using the Gross-Pitaevskii equation and compare the simulation results to the model predictions. 
\end{abstract}

\maketitle
\section{Introduction}
Hydrodynamic phenomena in Bose-Einstein condensates (BEC) have
recently become of interest, in particular, studies of the
Kelvin-Helmholtz (KH) and the Rayleigh-Taylor (RT) instabilities
\cite{Blaauwgeers2002, Volovik2002, Sasaki2009, Takeuchi2010} have
been undertaken. As quantum gases, BECs introduce new types of terms into the hydrodynamic equations, which may then lead to nonclassical behavior. For example, recent studies of  shock waves in BECs and quantum plasmas demonstrated qualitatively different structures of the quantum shocks in comparison with the classic gas- and plasma dynamics  \cite{Hoefer-et-al-2006,Hoefer-et-al-2008,Bychkov-et-al-2008}.
For the KH and RT instabilities, it has been seen that the quantum effects led to an upper cutoff for the unstable wavenumbers \cite{Sasaki2009,Takeuchi2010}. In addition, BECs allow for wide tunability of the experimental parameters, and thus they are excellent systems for exploring large parameter regimes and quantitatively testing theoretical predictions.

Here, we will consider the
Richmyer-Meshkov (RM) instability, another
fundamental hydrodynamic instability in BEC. The classical RM instability in
gases is closely related to the RT instability. The RT instability
develops when a heavy gas is supported by a light gas in a
gravitational field, while the RM instability is produced by a
pulsed acceleration instead of a constant one \cite{Richmyer1960,
Meshkov1969, Dimonte2010}. The acceleration pulse typically occurs
due to a shock wave hitting an interface between two gases of
different density. In the classical case, the RM and RT
instabilities demonstrate a qualitatively similar pattern of
spikes and bubbles with mushroom-structure, though the RM
instability develops slower, both at the linear and non-linear
stages. A quantum counterpart of the RT instability in
BEC was recently demonstrated in \cite{Sasaki2009}. Sasaki et al.
\cite{Sasaki2009} imitated the real gravity of the classical case
by the gradient of a magnetic field in the system of two coupled
BECs with different spins, e.g. for two hyperfine states of
$^{87}\textrm{Rb}$. Numerical simulations of the coupled Gross-Pitaevskii
equations demonstrated development of the mushroom structure in
agreement with the classical case. Stabilization of the
short-wavelength perturbations of the quantum RT instability was
observed due to surface tension between
the two BECs.

The RM instability in the system of a two-component BEC subject to
a magnetic pulse looks like a natural extension of the RT results.
Surprisingly, the RM problem of an interface with surface tension
has not been solved yet even in the classical case. Only recently,
Ref. \cite{Sohn2009} considered the problem but found that the
classical methods of the RT-RM theory (e.g. the Layzer model) fail
when surface tension influences the instability. Though brief and
qualitative, the discussion of Ref. \cite{Sohn2009} pointed out
that strong changes are expected in the instability development at
the nonlinear stage because of surface tension. In the following work, we show that dramatic changes happen even at the linear
stage of the instability, which were overlooked in
\cite{Sohn2009}. The main reason for these changes is that the
system possesses intrinsic dynamical properties determined by the
capillary waves, and this is qualitatively different from the
classical RM instability at an inert interface. In that sense, the
present case resembles the RM instability at a flame front studied
in Refs. \cite{Bychkov1998, Travnikov1999}, for which the
intrinsic flame evolution dominates asymptotically over the
temporary effects of a passing shock.

    In the present paper we investigate the magnetically induced RM instability in a two-component BEC. We construct and study analytical models describing the instability development both at the linear and nonlinear stages. The models indicate new features of the instability: influence of quantum capillary waves and separation of droplets, which are qualitatively different from the classical case. We perform numerical simulations of the instability in a trapped BEC using the Gross-Pitaevskii equation and compare the simulation results to the model predictions.

\section{Analytical models}
The dispersion relation for
interface waves in
a BEC
consisting of two components with different spins in a magnetic field gradient $B'$  may be presented as \cite{Sasaki2009}
\begin{equation}
-\omega^2 = \mu_B B'k/2m - \omega_c^2,
\label{dispersion}
\end{equation}
where $\omega$ is the perturbation frequency, $k=2\pi/\lambda$ is the wave number, $\lambda$ is the wavelength of the perturbation, $\mu_B$  is the Bohr magneton, and $m$ is the atomic mass. In the case of zero
magnetic field, the dispersion relation describes capillary waves with frequency
\begin{equation}
\omega_c^2 = \sigma k^3 /2nm,
\label{omegac}
\end{equation}
 where $\sigma$ is the surface tension and $n$ is the condensate concentration. A positive magnetic gradient in the dispersion relation leads to the magnetic counterpart of the RT instability with $\textrm{Im} (\omega)>0$, $\textrm{Re} (\omega)=0$ for sufficiently long perturbation wavelength. In the present work we are interested in the RM instability, which develops in the case of an impulsive magnetic field $\mu_B B'/2m = U\delta(t)$, where the amplitude $U$ has the dimension of velocity. For this case, the dispersion relation Eq. (\ref{dispersion}) should be re-written as a time-dependent differential equation with respect to the interface position $z=f(t)\exp(ikx)$, as
\begin{equation}
\frac{d^2f}{dt^2}= [U\delta(t)k - \omega_c^2]f.
\label{RM-linear}
\end{equation}
The original analysis by Richtmyer and Meshkov \cite{Richmyer1960, Meshkov1969, Dimonte2010} considered the case of zero surface tension $\sigma=0$, which gives the capillary wave frequency $\omega_c=0$. In the classical geometry of the RM instability, taking the initial condition $f=f_0$ at $t=0$, one can integrate Eq. (\ref{RM-linear}) to find linear growth of the perturbation, $f=Ukf_0t+f_0$. Surface tension modifies the solution to Eq. (\ref{RM-linear}) drastically. Solving Eq. (\ref{RM-linear}) for $t<0$ we find capillary waves
\begin{equation}
f=f_r\exp(ikx-i\omega_ct)+f_l\exp(ikx+i\omega_ct),
\label{capillary}
\end{equation}
where labels $r$ and $l$ denote waves propagating to the right and to the left, respectively.  The capillary waves determine the initial conditions for $f_-$ and $(df/dt)_-$ at the interface just before the magnetic pulse at $t=-0$. Conditions just after the pulse, at $t=+0$ , are obtained by integrating Eq. (\ref{RM-linear}) as
\begin{equation}
\left[ f \right]^{+}_{-}=0,  \quad \left[ df/dt \right]^{+}_{-}=Ukf_{-},
\label{initial conditions}
\end{equation}
where $\left[ ... \right]^{+}_{-}$ designates changes of the respective value during the pulse. Still, the solution to
 Eq. (\ref{RM-linear}) after the pulse at $t>0$ is also a superposition of the capillary waves of Eq. (\ref{capillary}) with new amplitudes $F_r$, and $F_l$. The new amplitudes are related to the initial amplitudes using Eq. (\ref{initial conditions}), and are found to be
\begin{equation}
F_{r,l} = f_{r,l} \pm i \frac{Uk}{2\omega_c}(f_r + f_l).
\label{new amplitudes}
\end{equation}
Thus, instead of the linear perturbation growth inherent to the classical case we obtain energy pumping and redistribution in the intrinsic capillary waves of the BEC. In the limit of a very strong pulse, $Uk/\omega_c \gg 1$, we observe the tendency of a standing wave to form out of an initial arbitrary wave distribution. If we take initial perturbations in the form of a standing wave as well with $f_r = f_l^* = (1/2)f_s\exp(i\alpha)$, $f_s$ and $\alpha$ being the amplitude and the phase, then Eq. (\ref{new amplitudes}) yields the new standing wave amplitude $F_s$ as
\begin{equation}
F_s = f_s \sqrt{\cos^2\alpha + \left(\sin\alpha + \frac{Uk}{\omega_c}\cos\alpha\right)^2 }.
\end{equation}
The transformation of magnetic energy to the hydrodynamic one is the
most effective for the magnetic pulse acting on BEC when the standing capillary wave
is at it's maximum, $\alpha=0$ , and this gives
\begin{equation}
 F_s = f_s \sqrt{1+U^2k^2/\omega_c^2}.
\label{relative_growth}
\end{equation}
 In the opposite case, with $\alpha=\pi/2$, the pulse does not produce any effect on the interface at all.
 The combination $Uk/\omega_c$ plays the role of the Weber number, $\textrm{We}=U^2k^2/\omega_c^2=2nmU^2/\sigma k$.

\begin{figure}
\begin{center}
\includegraphics[width=3in, keepaspectratio]{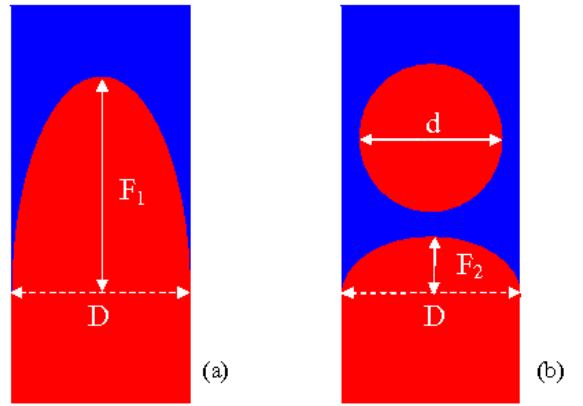}
\caption{(color online) Detachment of  droplets in the RM instability}
\label{drop_separation}
\end{center}
\end{figure}

    The linear analysis holds as long as the perturbation amplitudes are sufficiently small, with $kF_s \ll 1$.
    At the nonlinear stage, the classical RM instability leads to bubbles and spikes, which resemble
    qualitatively the nonlinear stage of the RT instability. The bubble velocity tends asymptotically to a
    constant for the RT geometry, and it slows down for the RM configuration according to $1/kt$ \cite{Dimonte2010}.
    For the case of a BEC we expect a qualitatively different behavior of the instability due to two physical
    mechanisms: 1) The perturbation growth is finally stopped and pushed back by surface tension;
    2) An elongated perturbation finger becomes energetically unstable with respect to detachment of droplets.
    The first mechanism may be understood already from the above linear analysis. To investigate the second mechanism,
    we study the system illustrated in Fig. \ref{drop_separation}. We compare the circumference of the perturbation
    finger of amplitude $F_1$ to a similar finger of smaller amplitude $F_2$ plus a circular droplet of diameter $d$
    for a fixed total surface area of the red
part of the
figure. Taking, for example, a sinusoidal shape of the perturbations,
    we obtain that droplet detachment provides a gain in the surface energy for the initial amplitude exceeding the
    critical value $F_1/D\approx 2.70$, which corresponds
    to $F_2/D\approx 0.28$ and $d/D\approx 0.70$. Taking an axisymmetric counterpart of Fig. 1
    we obtain similar estimates for the critical amplitudes $F_1/D\approx 1.42$, $F_2/D\approx 0.50$
     with the radius of a spherical droplet $d/D \approx 0.33$.
Of course, such a model provides mainly qualitative understanding of the nonlinear stage of the RM
instability in BEC, and other processes are expected to make the interface dynamics more complicated.
For example, the secondary KH instability develops at the interface between two components gliding along
each other at different velocities \cite{Blaauwgeers2002, Volovik2002, Sasaki2009, Takeuchi2010}.
Because of the secondary KH instability the perturbation fingers may acquire a mushroom-shape, and
thus spread beyond the limits of the channel sketched in Fig. \ref{drop_separation}.

\section{Numerical simulations}\label{Formalism}

We take our system to be described by the two coupled Gross-Pitaevskii equations (GPEs)
\begin{align}
i\hbar\frac{\partial\psi_1}{\partial t}&=&\left\{-\frac{\hbar^2}{2m_1}\nabla^2 + V_1(\mathbf{r}) + g_{11}|\psi_1|^2 + g_{12}|\psi_2|^2\right\} \psi_1 ,\nonumber\\
i\hbar\frac{\partial\psi_2}{\partial t}&=&\left\{-\frac{\hbar^2}{2m_2}\nabla^2 + V_2(\mathbf{r}) + g_{22}|\psi_2|^2 + g_{12}|\psi_1|^2\right\} \psi_2 , \label{GPE}
\end{align}
where $\psi_j$ is the macroscopic wavefunction, $V_j$ is the harmonic trapping potential on component $j$, and the interaction parameter is given as
\begin{equation}
g_{jj'}^{3D} = \frac{2\pi \hbar^2 a_{jj'}}{m_{jj'}},
\end{equation}
with $a_{jj'}$ the s-wave scattering length and $m_{jj'}$ the reduced mass.


For our calculations we employ a two dimensional (2D) system  with a modified interaction parameter, $g_{jj'}^{2D}$, so that our 2D simulation can be related to a 3D tight pancake trap. Orienting our system so the tight direction is in $z$, our interaction parameter is modified as follows \cite{Pethick}:
\begin{equation}
g_{jj'}^{2D} = \frac{\sqrt{2\pi} \hbar^2 a_{jj'}}{m_{jj'}a_z},
\end{equation}
where $a_z$ is the harmonic oscillator length in the $z$ direction. We choose our two components to be the two hyperfine states of $^{87}$Rb, that is $|F,m_F\rangle = |1,1\rangle$ and $|1,-1\rangle$. These have scattering lengths \cite{vanKempen2002} $a_{11}=a_{22}=100.4a_B$ and $a_{12}=101.3a_B$ where $a_B$ is the Bohr radius.

We begin with a system of $~3.2\times10^7$ atoms of $^{87}$Rb, equally split between the two hyperfine states. We use the trap geometry $\omega_x=\omega_y=2\pi\times100$Hz and $\omega_z=2\pi\times5$kHz.
We solve numerically the coupled GP equations to find the ground state of the system. The two components are confined to $y>0$ and $y<0$ respectively. Our numerical investigation into the RM instability is arranged as follows.
 First, we induce capillary waves on the interface between the condensates and study their evolution, this is detailed in Sec. \ref{sec:cap_waves}. The standing capillary waves are then used as the initial state for the RM instability. We trigger the RM instability at the interface, using a pulsed magentic field in Sec. \ref{sec:RM_inst} and investigate the different regimes of the instability. We then make a comparison between our analytical models and numerical results, and explore reasons for any discrepancy in Sec. \ref{sec:comp_ana}.

\subsection{Capillary waves}\label{sec:cap_waves}

To
 form capillary waves at the interface of the two condensates, we introduce a very small
 sinusoidal perturbation of $\lambda=$9$\mu$m to each
component along the interface. A steady magnetic gradient of
$B'=1.78$G/cm is added to the system, directed so that the
condensates are pushed toward one another. This results in the
amplification of the perturbation, and the eventual formation of
the RT instability \cite{Sasaki2009}. When the
waveform on the surface has reached suitable amplitude, we switch
off the magnetic gradient and allow the capillary wave to
oscillate freely.

\begin{figure}
\begin{center}
\includegraphics[width=3in, keepaspectratio]{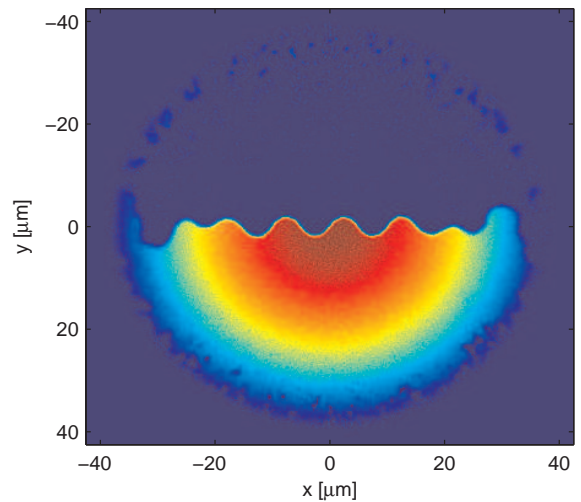}
\caption{(color online) Density of the lower component, showing the capillary waves on the interface at the point of maximum amplitude.}\label{cap_density}
\end{center}
\end{figure}

Surface tension and hence the frequency of oscillations depends on
the density at the interface as $\sigma \propto n^{3/2}$,
$\omega_{c} \propto n^{1/4}$, with a maximum at the peak density, and reducing to near zero at the edges of the system. This
introduces finite size effects into our system, and for this
reason, any quantitative results will focus on the system behavior
at peak density. We measure the frequency of oscillation at this
point to be $\omega_{c}=246s^{-1}$. An example of the capillary
wave in the system at peak amplitude of oscillation is shown in
Fig. \ref{cap_density}. While the waves in the center of the
system are uniform, we observe irregularities on the edge of the
cloud. The irregularities develop due to the variations of the
oscillation frequency  across the wavefront, which lead to a
continuous phase shift of the capillary waves at the center and on
the edges.

A further effect of the reduced surface tension and intercomponent
repulsion on the edges of the cloud is the 'pass-through' of
condensate from one side of the system to the other.
Interpenetration of the condensates has been observed already in
simulations \cite{Sasaki2009}. Our numerical results show the same
effect, though in a reduced form as compared to that in
\cite{Sasaki2009}: our choice of initial perturbations allows the
interface instability to develop faster, and so there is less time
for condensate pass-through to occur. We use the configuration
shown in Fig. \ref{cap_density} as the initial state for studying
the RM instability in the following sections.

\subsection{RM instability for different pulse strength}\label{sec:RM_inst}

We apply a magnetic pulse of the form $B'\propto
\exp\left(-t^2/\tau^{2}\right)$ to the system, where $\tau=0.02$
ms controls the duration of the pulse. We note that while the
field is not a delta function as in our analytical model, the
pulse duration is much less than the characteristic time scale of
the capillary wave oscillation, $\omega_{c}\tau \ll 1$. The
magnetic "shock velocity" of the analytical model is calculated
as
\begin{equation}
U=\sqrt{\pi}\frac{\mu_{B}B'\tau}{2m}.
\end{equation}
We administer the shock to the initial state shown in Fig.
\ref{cap_density}. The behavior of the system following the pulse
depends on the size of the magnetic field gradient $B'$. In
agreement with our theoretical predictions, we see a wide range of
behavioral regimes, comprising of both linear and nonlinear
dynamics. We can further split the nonlinear behavior into three
distinct regimes: simple nonlinear, nonlinear with detachment, and
turbulent nonlinear. We detail the defining features of these
regimes in the remainder of this section.


In the linear phase typical for rather weak magnetic pulses with
$U \leq 3\times 10^{-4} \textrm{ms}^{-1}$, $\textrm{We} \leq 0.70$, the pulse results in
amplification of the waveform on the interface without qualitative
modifications of its' shape. After a short phase of growth, the
waveform then reduces in height, and eventually returns to
capillary oscillations. Throughout the entirety of the evolution
of the system the wave at the interface remains sinusoidal.

\begin{figure}
\begin{center}
\includegraphics[width=3in, keepaspectratio]{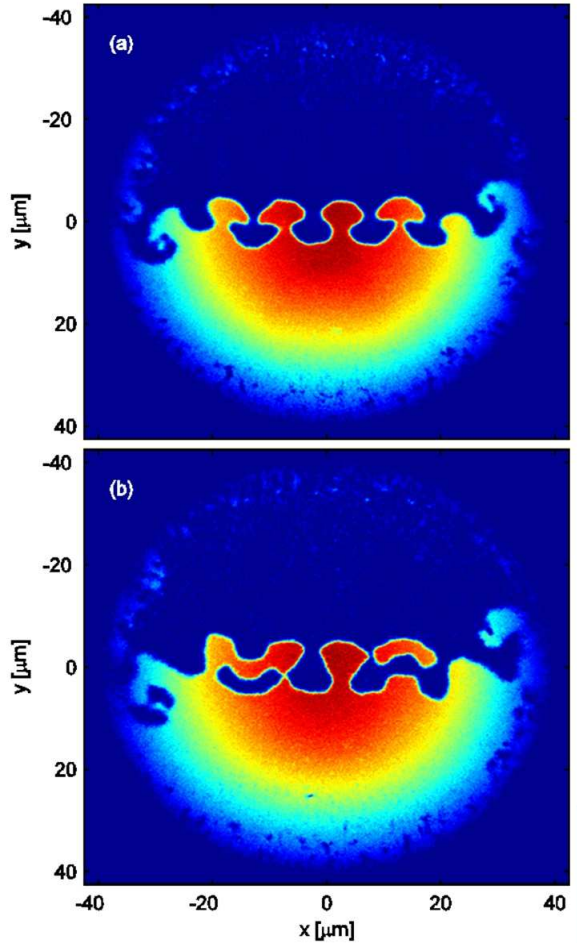}
\caption{(color online) Cloud density of one component for times 6.4 and 12.7ms after shock size $U=0.75\times10^{-3}\textrm{ms}^{-1}$}\label{nonlinear_simple}
\end{center}
\end{figure}

To demonstrate the simple nonlinear phase, in Fig.
\ref{nonlinear_simple} we show the system at times 6.4ms and
12.7ms after a pulse of strength $0.75\times10^{-3}\textrm{ms}^{-1}$, which
corresponds to the Weber number $\textrm{We}=4.4$. After the pulse, the
sinusoidal waveform grows at a rate proportional to the pulse
amplitude, in agreement with the analytical model, Eq.
(\ref{initial conditions}). As the perturbation humps grow beyond
the limits of linear dynamics $kf \ll 1$, they become distorted, and
then eventually 'pinch' in to form bubbles, as can be seen in Fig.
\ref{nonlinear_simple} A. At the later time in Fig.
\ref{nonlinear_simple} B we can see that some of these bubbles (on
the edges of the cloud) have detached, while others in the middle
have remained connected to the main body of the condensate. The
difference in behavior happens due to variations of surface
tension and local Weber number across the interface. Because of
the lower surface tension the magnetic pulse causes greater
disturbance on the cloud edges making the flow pattern similar to
the classical RM instability \cite{Richmyer1960, Meshkov1969,
Dimonte2010}. At the very edges of the cloud in Fig.
\ref{nonlinear_simple}(a), where surface tension is almost zero, the
growing perturbations acquire the shape of a half of a "mushroom".
The clearly visible vortex spiral at the side of the "mushroom"
cap is a characteristic feature of the classical RM instability
for gases of comparable density with zero surface tension. The
vortex develops as a result of the secondary KH instability of two
gases (fluids, condensates) gliding along each other because of
the primary RM instability. In our case the condensates are of the
same density, which creates optimal conditions for the secondary
KH instability at the cloud edges. Qualitatively different
dynamics of the condensate at the trap center and at the edges
may illustrate the difference between the "classical" and quantum RM
instability.

\begin{figure}
\begin{center}
\includegraphics[width=3in, keepaspectratio]{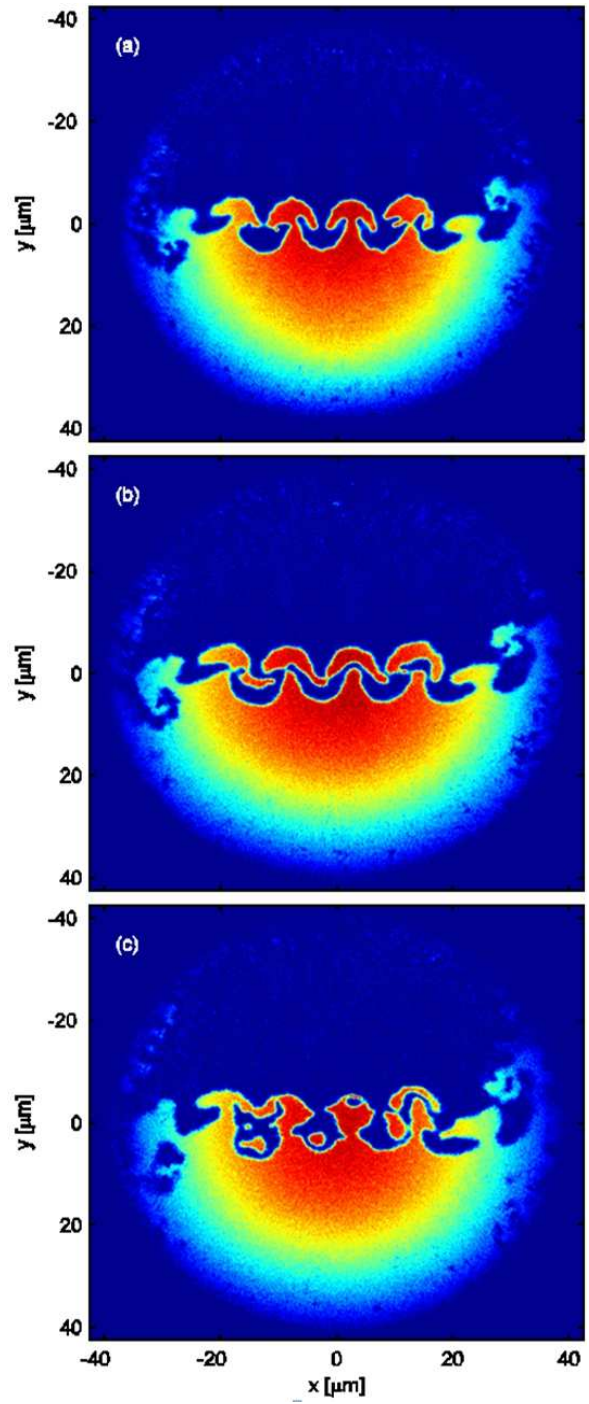}
\caption{(color online) Cloud density of one component for times 5.7, 8.0 and 12.7ms after shock size $U=1.05\times10^{-3}\textrm{ms}^{-1}$} \label{nonlinear_detached}
\end{center}
\end{figure}

\begin{figure}
\begin{center}
\includegraphics[width=3.5in, keepaspectratio]{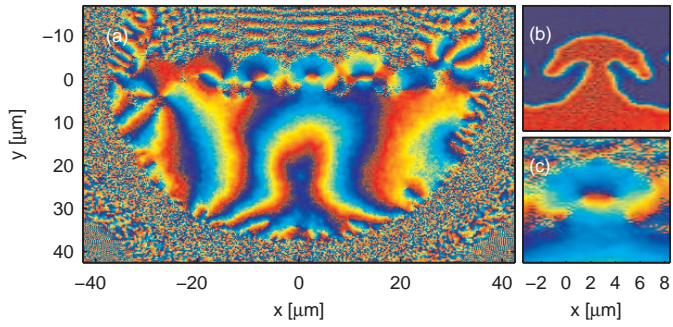}
\caption{(color online) Phase of portion of a cloud, 5.7ms after shock of size $U=1.05\times10^{-3}\textrm{ms}^{-1}$ (a), showing density of a single finger (b), and phase of a single finger (c).}\label{vortex}
\end{center}
\end{figure}

For a stronger magnetic pulse $U \geq 8\times 10^{-4} \textrm{ms}^{-1}$, $\textrm{We}
\geq 5.0$, the nonlinear phase is accompanied by detachment of
droplets all over the interface. We present the system at times
5.7ms, 8.0ms and 12.7ms after a pulse of strength $U=1.05\times10^{-3}\textrm{ms}^{-1}$,
$\textrm{We} = 8.6$ in Fig. \ref{nonlinear_detached}, which demonstrates some
new features in the system evolution. In Fig
\ref{nonlinear_detached}(a) we find that instead of simple bubble
formation we get mushroom-shaped caps reminiscent of RT
instability. It is interesting to note the different shape of
these mushrooms, as compared to those produced due to the RT
instability \cite{Sasaki2009}: for the RM instability, the
mushroom caps have less spiralled density under the caps. We
conclude that the surface tension has a much larger effect for a
pulse acceleration, inhibiting the growth of the secondary KH
instability. The respective phase pattern of the system at
$t=5.7$ms is shown in Fig. \ref{vortex}. We can see that there are
indeed vortex pairs forming on the caps in Fig.
\ref{nonlinear_detached}(a). These vortices then evolve in a
complicated manner, giving rise to results seen for example in
Fig. \ref{nonlinear_detached}(b),(c). Another important feature of the
phase pattern Fig. \ref{vortex} is the large-scale acoustic
oscillations of the trapped condensate as a whole (the "breathing"
mode). As we demonstrate below in Section \ref{sec:comp_ana},
the acoustic oscillations influence development of the RM
instability noticeably, leading to quantitative deviation of the
numerical results from the analytical model of an incompressible
flow without confinement. The distinctive feature of the quantum
RM instability for sufficiently strong magnetic pulses is
detachment of the mushroom caps  from the main cloud shown in Fig.
\ref{nonlinear_detached}(b). The detachment occurs when the distance
from peak to trough is approximately twice the effective finger width, which correlates
reasonably well with the nonlinear analytical model, taking into
account limitations of the model. In the present case the magnetic pulse is not so strong, and the droplets are reattached back to the main cloud after a short time interval. The process of reattachment is accompanied by formation of
small-sized droplets and pockets in the condensate components.

\begin{figure}
\begin{center}
\includegraphics[width=3in, keepaspectratio]{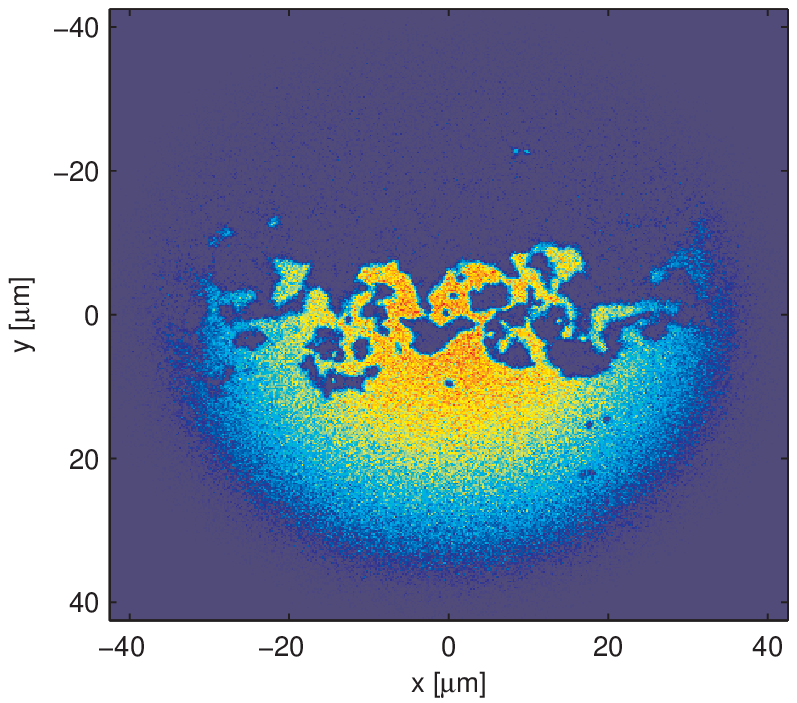}
\caption{(color online) Cloud density of one component at time 12.7ms after shock size $U=2.1\times10^{-3}\textrm{ms}^{-1}$} \label{nonlinear_turb}
\end{center}
\end{figure}

Finally, the instability effects become quite strong for an
intense magnetic pulse leading to turbulent interface dynamics. In
Fig. \ref{nonlinear_turb} we present an example of the highly
turbulent system 12.7ms after a pulse of strength
$U=2.1\times10^{-3}\textrm{ms}^{-1}$ corresponding to the Weber number $\textrm{We} = 34.5$.
This regime is characterized by the formation of multiple vortex
pairs, and by highly complex evolution of the interface. In Fig.
\ref{nonlinear_turb}, one can hardly identify separate humps,
fingers or mushrooms clearly visible in the previous instability
regimes. Instead, the interface takes the form of a mixing layer
resembling the RT cascade of extreme strength
\cite{Cabot-Cook-2006,Bofetta-et-al-2010}.

\subsection{Comparison with analytical results}\label{sec:comp_ana}

\begin{figure}
\begin{center}
\includegraphics[width=3.5in, keepaspectratio]{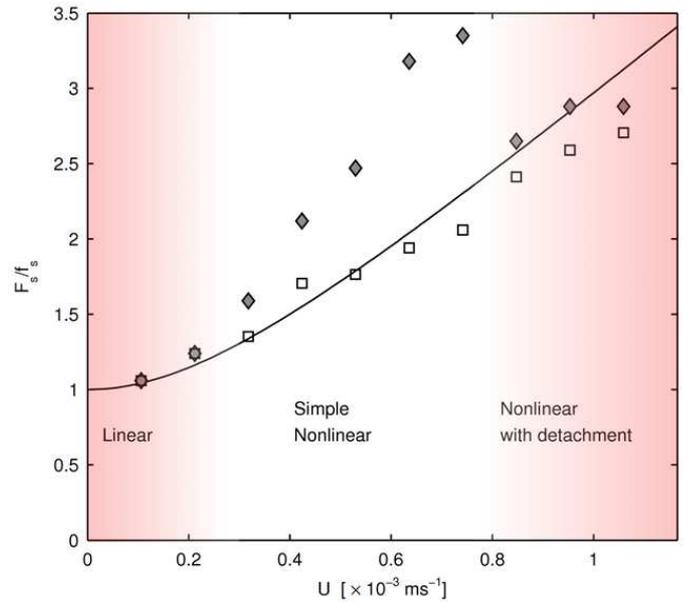}
\caption{(color online) Plot of relative fringe growth vs shock magnitude, U. Analytic prediction (solid line), and numerical simulation for total finger growth (filled diamonds) and growth until first levelling off (open squares), see text.} \label{Many_U}
\end{center}
\end{figure}

We now compare our simulation results to the analytical model.
Figure \ref{Many_U} shows the analytical predictions of Eq.
(\ref{relative_growth}) for the increase in the hump amplitude
because of the magnetic pulse (presented by the solid line) and
the respective numerical data.
 In the simulations, we measure the perturbation amplitude
in the very center of the system just before the pulse, and  the
maximal amplitude after the pulse resulting from the RM
instability; their ratio is shown in Fig. \ref{Many_U} by filled
diamonds. Gradient shading in the background indicates different
regimes of the instability.

At low pulse strength, in the linear regime, we see good agreement
between simulation and theory. As the pulse strength is increased,
in the simple nonlinear regime, the simulation results deviate
from the theoretical predictions, demonstrating noticeably
stronger perturbation growth. At very high pulse strength, in the
nonlinear regime with detachment of droplets, we see that the
theory and simulations once again come to agreement.

The deviation of the numerical data from the theoretical
predictions is related to the finite size of the system and to the
acoustic oscillations of the trapped condensate. As we apply a
magnetic pulse to the system, the condensate gets compressed in
the direction of the magnetic force (along the y-axis), and it expands
again after the pulse. Compression produces density peak close to
the interface, which propagates away in the form of a weak shock.
When reflected from the system boundaries, the shock excites acoustic modes in the condensate. The large-scale semi-circles in the phase pattern of Fig. \ref{vortex}(a)
correspond to the acoustic modes. We demonstrate below that the
acoustic oscillations pump extra energy in the development of the
RM instability, thus leading to additional growth of perturbations
in comparison with predictions of the analytical model.

\begin{figure}
\begin{center}
\includegraphics[width=3.5in, keepaspectratio]{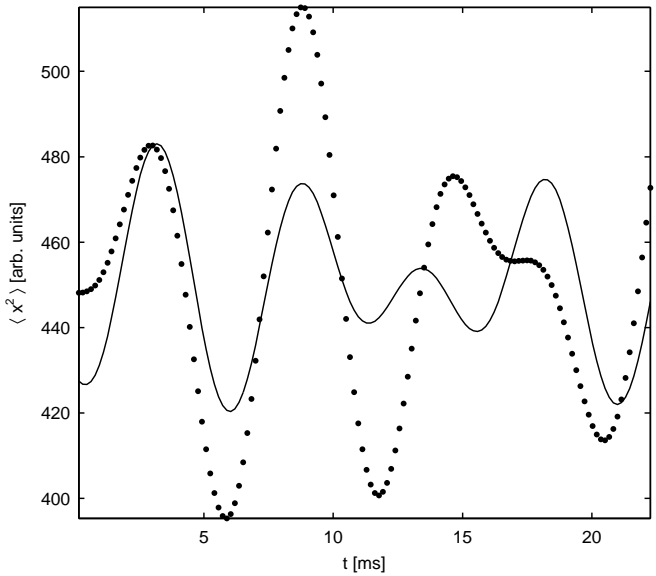}
\caption{Width of cloud in
x direction (parallel to the interface) as a function of time for perturbed (solid line) and unperturbed (dotted line) clouds.} \label{excite}
\end{center}
\end{figure}

\begin{figure}
\begin{center}
\includegraphics[width=3.5in, keepaspectratio]{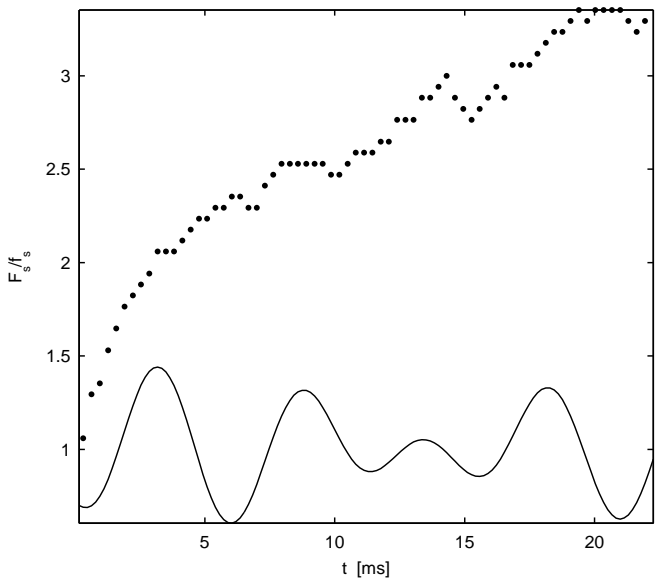}
\caption{Relative growth of fringe as function of time for pulse $U=0.75\times10^{-3}\textrm{ms}^{-1}$ (filled circles) and width of the cloud, $\langle x^2 \rangle$ (solid line, arbitrary scale in y)} \label{70_grow}
\end{center}
\end{figure}

To study the excitations, we calculate
the mean-square width in the $x$ direction, $\langle x^2\rangle$,
as a function of time.
Evidently, the acoustic mode produces time oscillations of the
cloud width. To elucidate the interaction of the acoustic mode and
the RM instability, we perform two simulation runs: 1) First, we
apply magnetic pulse to the system with initially flat interface,
thus eliminating the RM instability; 2) Second, we study a similar
pulse acting on a system with perturbed interface and with the RM instability. Both cases are
studied in response to a pulse of $U=0.75\times10^{-3}\textrm{ms}^{-1}$.
The respective acoustic oscillations  are presented in Fig.
\ref{excite}. We see that for a flat interface (dotted line), the acoustic oscillations demonstrate a noticeably larger
amplitude, also, the beat in the frequency of width oscillation seen around 15ms signals the presence of more than one mode. In the case of a perturbed interface, the oscillation
amplitude is markedly lower, which indicates that some of the
oscillation energy is dispersed through the instability on
interface.
We can also compare  time dependence for the RM perturbation
growth and the acoustic oscillations, see Fig. \ref{70_grow}. As
we can see, the rapid initial growth of the interface perturbations
(dotted line) is followed by a leveling off of growth at around
$t=3$ms. The subsequent periods of perturbation growth and
slowdown correlate quite well with acoustic oscillations shown by
the solid line in Fig. \ref{70_grow}.
In this respect we can see why  the growth of RM perturbations in
the numerical modeling is noticeably larger than that predicted
by theory.  It is likely that additional perturbation growth  is
provided by the energy of acoustic oscillation pumping the
instability. In contrast, the theoretical model assumes that the shock wave travels away from the interface region and never returns.
The effect of extra pumping may be compared to the
classical RM instability in gases under confinement
\cite{Collins-Jakobs-2002,Schilling-et-al-2007}.
 In the geometry of Refs. \cite{Collins-Jakobs-2002,Schilling-et-al-2007},  secondary shocks/acoustic waves reflected from the bounding walls produce
 secondary RM instabilities and additional corrugation of the
 unstable interface. Pumping of the RM/RT instabilities by
 acoustic waves in confined combustion systems have been also observed in
 \cite{Petchenko-et-al-2006,Petchenko-et-al-2007}.
 To minimize the influence of acoustic pumping on the RM growth in
 our simulations, we measure the perturbation amplitude at the
 instant of the first leveling off, which corresponds to the
 first maximum in the acoustic mode. The respective amplitudes
 shown on Fig. \ref{Many_U} by open squares, demonstrate a good
 agreement with the theoretical predictions of Eq.
(\ref{relative_growth}) for all regimes of the instability. We also
point out that the difference between two ways of measuring the
final amplitude is minor in the nonlinear regime with the detachment
of droplets. It is interesting to note that  droplets typically
detach close to the maximum of the acoustic mode, during the
second contraction of the quadrupole excitations. Once detached, the
droplets are no longer subject to the extra instability pumped by
the oscillations of the cloud, and so they stop growing. At very
large times, once the droplets reattach back to the main
condensate cloud, we note that they can begin to grow again.

\section{Conclusions and outlook}\label{Conclusion}
Thus, in the present paper we have demonstrated the possibility of the RM
instability in a quantum system of a two-component BEC consisting
of hyperfine states of $^{87}\textrm{Rb}$ with spins pointing in
the opposite directions. The instability is triggered by a
magnetic pulse pushing the BEC components towards each other. We
develop the analytical models of the instability at the linear and
nonlinear stages, and solve coupled GP equations numerically to
study the RM instability. Both theory and simulations indicate new
features of the instability that are different from the classical case and
related to quantum surface tension and capillary waves. We obtain
different regimes of the instability depending on the strength of
the magnetic pulse: the linear regime, the simple nonlinear
regime, the nonlinear regime with detachment of droplets and the
regime of turbulent distortion and mixing of the unstable
interface. Taking into account the influence of acoustic oscillations
of the trapped condensate, we find good quantitative agreement of
our analytical model and  numerical simulations.

We point out that experimental realization of the RM instability
requires a capillary wave of non-negligible amplitude as an initial condition. We have shown in this paper how such an initial condition can be created in a BEC by applying a weaker magnetic field over a longer time, thus making use of the RT instability. In a realistic experiment, the initial state will be random rather than containing a sinusoidal interface perturbation and thus the RT field will excite a spectrum of wavelengths rather than a single one; nevertheless, the qualitative features will be unchanged. A single-wavelength interface wave could conceivably be excited by applying a set of localized optical potentials along the interface or possibly combining an optical potential with a mask to give it the desired spatial shape.




\end{document}